\documentclass[preprint,5p]{elsarticle}

\usepackage{amsmath}
\usepackage[english]{babel}
\usepackage[utf8]{inputenc}
\usepackage[T1]{fontenc}
\usepackage{accents}
\usepackage{graphicx}
\usepackage{subfigure}
\usepackage{amssymb}
\usepackage{xcolor}
\usepackage[table,xcdraw]{}
\usepackage[normalem]{ulem}
\usepackage{xspace}
\usepackage{units}
\usepackage{multirow}
\biboptions{sort&compress}
\usepackage{hyperref}

\def\GeV{\nobreak\,\mbox{GeV}}

\newcommand{\RAA}{R$_{\text{AA}}$\xspace}
\newcommand{\pt}{p$_\perp$\xspace}
\newcommand{\vtwo}{$v_2$\xspace}
\newcommand{\avgt}{$\left< \text{n}_{\text{int}} \right> \cdot (\text{DM})^2$\xspace}

\begin{document}

\title{How many interactions does it take to modify a jet?}
\author[lund]{Chiara Le Roux}
\author[lip,ist]{José Guilherme Milhano}
\author[lund]{Korinna Zapp}

\address[lund]{Dept. of Physics, Lund University, Sölvegatan 14A, S22362 Lund, Sweden}
\address[lip]{LIP, Avenida Prof. Gama Pinto, 2 P-1649-003 Lisboa, Portugal}
\address[ist]{Departamento de Física, Instituto Superior T\'ecnico (IST), Universidade de Lisboa, Av. Rovisco Pais 1, P-1049-001 Lisboa,Portugal}

\begin{abstract}

It is a continued open question how there can be an azimuthal anisotropy of high $p_\perp$ particles quantified by a sizable $v_2$ in p+Pb collisions when, at the same time, the nuclear modification factor $R_\text{AA}$ is consistent with unity. We address this puzzle within the framework of the jet quenching model \textsc{Jewel}. In the absence of reliable medium models for small collision systems we use the number of scatterings per parton times the squared Debye mass to characterise the strength of medium modifications. Working with a simple brick medium model we show that, for small systems and not too strong modifications, $R_\text{AA}$ and $v_2$ approximately scale with this quantity. We find that a comparatively large number of scatterings is needed to generate measurable jet quenching. Our results indicate that the $R_\text{AA}$ corresponding to the observed $v_2$ could fall within the experimental uncertainty. Thus, while there is currently no contradiction with the measurements, our results indicate that $v_2$ and $R_\text{AA}$ go hand-in-hand. We also discuss departures from scaling, in particular, due to sizable inelastic energy loss. 

\end{abstract}

\maketitle

\section{Introduction}

In the field of relativistic heavy ion collisions, the idea that high energy partons travelling through a medium will lose energy due to successive interactions with the medium constituents has long been established. This was done from the theoretical point of view \cite{bjEloss,BDMPS} (for reviews see~\cite{Wiedemann:2009sh,Majumder:2010qh,Mehtar-Tani:2013pia,Qin:2015srf,Apolinario:2022vzg}) and subsequently confirmed experimentally \cite{expQuenching1,expQuenching2} (for reviews see~\cite{Connors:2017ptx,Cunqueiro:2021wls}). The experimental confirmation was carried out via the measurement of nuclear modification factors, which  compare the yields of high \pt particles or jets in heavy ion collisions to those in pp collisions, where no medium was expected to be formed. These measurements were then repeated in small collision systems such as d+Au and p+Pb and the absence of quenching was confirmed \cite{expSmallSys1,expSmallSys2,ALICE:2015umm,CMS:2015gcq,CMS:2016svx,ALICE:2021wct,ATLAS:2022iyq,ALICE:2023plt,ATLAS:2014cpa,Khachatryan:2015xaa,Adam:2016jfp,Adam:2016dau}. Such results motivated the interpretation of this suppression in the yields as a signature of the presence of a quark gluon plasma (QGP) phase in the early stages of relativistic heavy ion collisions.

As a means to investigate this newly discovered, short lived phase of matter, several other observables were proposed. One of them is the azimuthal momentum anisotropy, which is commonly characterised via the flow coefficients, $v_n$, appearing in the Fourier decomposition of the particle distribution
\begin{equation}
    \frac{dN}{d\phi} = \frac{1}{2} \left( 1 + 2\sum_{n} v_n \cos(2(\phi - \Psi_n)) \right)\,.
    \label{eq:fourier series}
\end{equation}
Here, $\Psi_n$ is the azimuthal angle of the n$^{th}$ symmetry plane. Since the overlap region of two colliding nuclei has a strong elliptical deformation, the corresponding elliptical term in the momentum distribution, $v_2$, is the most prominent one in heavy ion collisions. For $n=2$, $\Psi_2$ is the orientation of the short axis of the overlap region. Therefore, when all $v_n = 0$, the distribution is completely isotropic and, as it acquires an ellipsoidal shape, a non-zero \vtwo is observed.

Several experiments have measured the flow coefficients and found them to be non-zero in relativistic heavy ion collisions \cite{v21,v22,Wang:2024irg,STAR:2011ert,Korotkikh:2010iao}. In these collisions a non-vanishing $v_2$ of low transverse momentum particles is a consequence of the collective flow of the system. Because the pressure gradient driving the expansion is larger along the short axis of the overlap region than along the long axis, the particles are pushed out preferentially along the short axis. At high \pt, \vtwo is generated via the path length dependence of the energy loss suffered by the hard particle. It is thus also a consequence of the geometry of the system, but is not related to collective flow.

Surprisingly, sizable flow coefficients, in particular $v_2$, were also observed in small systems \cite{v2small1,v2small2}. These systems are believed to be too short lived to develop collective flow, but it has been shown in kinetic theory that, even with a low number of scatterings per particle, a sizable soft particle $v_2$ can be generated via the so-called escape mechanism~\cite{Kurkela:2018ygx,Ambrus:2021fej,He:2015hfa,Molnar:2019yam,Kurkela:2021ctp}. On the other hand, an explanation in terms of a hydrodynamic evolution of the system has also been put forward~\cite{Weller:2017tsr} (and criticised in~\cite{Zhou:2020pai}).

While viable explanations thus exist for the observation of non-vanishing $v_2$ of soft particles in small collision systems, the same is not true for the $v_2$ of high \pt particles found in such systems~\cite{ATLAS:2019vcm,ALICE:2022cwa,CMS-PAS-HIN-23-002}. According to our current understanding, that would have to be generated by anisotropic energy loss, but the absence of jet quenching contradicts such an interpretation. However, the question really remains a quantitative one, namely whether it is possible that a small amount of energy loss could generate a measurable high \pt \vtwo while not leading to a measurable $R_\text{AA}$ of hard particles or jets.

\smallskip

The present work aims to address precisely that question. To model the energy loss of hard partons, the \textsc{Jewel} event generator is used. Usually one must make assumptions of what a medium looks like and how it should expand, which, in turn, adds uncertainties about the assumptions in the specific the medium model used. In order to reduce the dependence on the medium model, we take a different approach and look specifically into how many jet-medium interactions one needs in order to see the aforementioned effects. In practice, we actually look at the magnitude of the average number of interactions per jet particle times the squared Debye mass of the medium \avgt, as will be discussed in the later sections. Therefore, we perform this study using a simplified medium model and then, knowing the \avgt needed to get effects of a given magnitude, we look into more realistic media. 

\section{JEWEL Monte Carlo Model}

The Monte Carlo model \textsc{Jewel}~\cite{Zapp:2012ak} simulates the QCD evolution of highly energetic partons produced in hard scattering processes in the presence of a background medi\-um. It is based on \textsc{Pythia}\,6.4~\cite{Sjostrand:2006za}, which provides the hard scattering matrix elements, initial state parton shower and hadronisation. \textsc{Jewel} has a virtuality ordered final state parton shower that is similar but not identical to the virtuality ordered parton shower in \text{Pythia}\,6. In vacuum this is an ordinary parton shower with the somewhat special feature that recoils from splittings are handled locally and in such a way that the algorithm never goes back to modify the kinematics of an earlier splitting. In the presence of a coloured medium, scattering off medium constituents can occur between the splittings generated by the parton shower. The scatterings are described by pQCD $t$-channel matrix elements regularised by the screening mass DM. If such a scattering is harder than the current parton shower scale it can reset the parton shower, re-starting it at the scale of the scattering. In this way, medium induced bremsstrahlung is effectively included and it is ensured that elastic and inelastic scattering occur with the leading-log correct relative rates. The Landau-Pomeranchuk-Migdal effect~\cite{Landau:1953gr,Migdal:1956tc} is included by allowing subsequent scatterings to act coherently if they fall within the formation time of the first splitting of a re-started parton shower. The momentum transfers are then added vectorially and the emission gets re-weighted with the inverse of the number of coherent momentum transfers. This procedure was shown to reproduce the BDMPS result in the eikonal limit~\cite{Zapp:2011ya}. Finally, re-starting the parton shower is only allowed when the first emission from the new shower has a shorter formation time than the current emission from the old shower. The medium partons recoiling from an interaction with a hard parton can be kept in the event to provide a simple model of medium response. The parton shower partons, and where applicable also the recoils, are hadronised with the \textsc{Pythia} string hadronisation. 
\textsc{Jewel} is largely agnostic about the background medium and it is therefore possible to interface with different medium models. However, it only simulates jets and medium response, but not the evolution of the bulk medium. Therefore, the events contain only particles that belong to the hard scattering and the parton showers or that have interacted with such a parton. For this study medium response is turned off in order to avoid complications in the interpretation of the results. 

\section{The small systems set-up}
\label{sec:small systems}

To address the puzzle of \vtwo and \RAA in small systems we have used \textsc{Jewel} with a \textit{brick}-like medium. This medium model consists of a collection of gluons distributed in an ellipsoidal region of space over which the temperature and density are uniform. The geometry of this region is defined by two input parameters: the length of the long axis of that ellipse (or sphere when the eccentricity is zero) and its eccentricity. The density and temperature can also be specified by input parameters. The Debye mass, which regularises the scattering cross section and controls the hardness of the interactions, is related to the temperature in the default \textsc{Jewel} setup. Here, we decouple it from the temperature and make it a free parameter allowing us to disentangle dependences.

All of the events (in medium or vacuum) used for the results in the upcoming sections are di-jet events at $\sqrt{s}=$ 5.02 TeV generated with $\hat p_\perp$ of the hard sctatering between 50 and \unit[500]{GeV}. In order to remove path length dependence (apart from the one coming from the geometry of the medium), we always generate the di-jets at the center of the brick. 

The two observables studied in the present work are \RAA of jets and \vtwo of high \pt particles. To compute them (using the Rivet framework~\cite{Bierlich:2019rhm}) we consider all the final state hadrons with \pt$ \geq 0.5$\GeV~within a pseudo rapidity range of $|\eta| < 2.8$. Then, for \RAA, we reconstruct jets using the anti-$k_\perp$ algorithm~\cite{Cacciari:2008gp} with a radius of $R=0.4$ and plot the \pt distributions of the reconstructed jets. To quantify the medium suppression, we integrate the spectra between 100 and \unit[400]{GeV} and take the ratio of the integrals in medium and in vacuum.

As for the \vtwo at high \pt, the default way to obtain it in small systems would be via correlations of high-\pt particles coming from the hard scattering with soft particles forming the background. This is done because the event plane in those systems is not well defined. However, the soft background is not available in \textsc{Jewel}. We thus make use of equation \ref{eq:fourier series} keeping in mind that, since the geometry of the brick is manually defined, the symmetry plane $\Psi_2$ is known. With that, and assuming all $v_{n>2} \approx 0$, we fit equation \ref{eq:fourier series}~to the normalised azimuthal angle distribution of all hadrons with \pt $\geq 2$\GeV~ and extract \vtwo as a parameter of the fit. This is done event by event and the \vtwo for a given set of parameters is the average for all the events in that set.

\begin{figure}[!ht]
    \centering
    \subfigure[]{
        \label{distribution-NB-mean}
        \includegraphics[width=.4\textwidth]{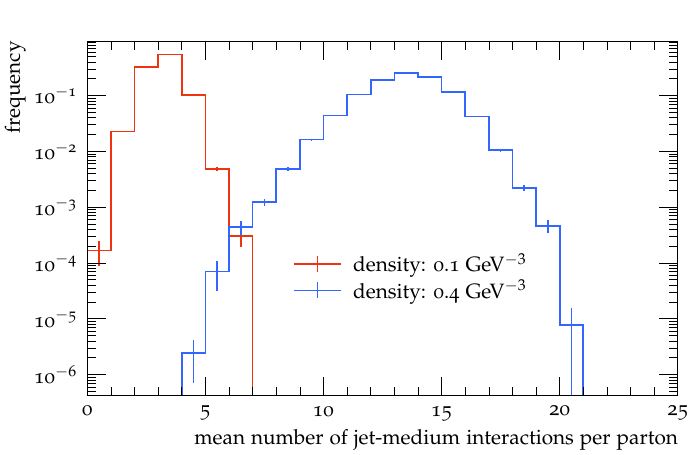}
    } \quad
    \subfigure[]{
        \label{distribution-NB-tot}
        \includegraphics[width=.4\textwidth]{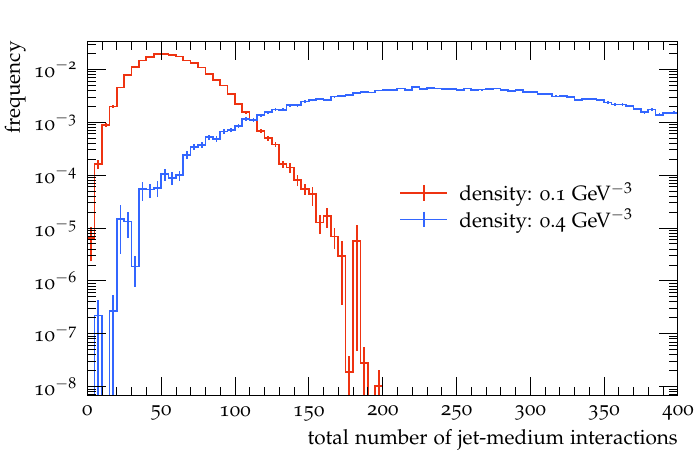}
    }
    \caption{Distributions of \subref{distribution-NB-mean} mean number of interactions per hard parton (see text for definition) in an event; and \subref{distribution-NB-tot} of total number of jet-medium interactions per d-jet for two different medium densities.}
    \label{fig:distributionNint}
\end{figure}

It is also important to clarify that, in order to understand the dependence of the observables of interest on the number of jet medium interactions, one must be careful not to introduce any biases related to the jet fragmentation pattern. Therefore, we calculate the observables of interest as a function of the \textit{average} number of interactions at a given density. In other words, the number of interactions is tuned by changing the density while keeping the temperature, the system size, and geometry fixed. In this way, we calculate the observables on a sample of jets with different fragmentation patterns and different number of interactions with the medium. As seen in figure~\ref{fig:distributionNint} the resulting distribution of the number of scatterings has a sizable width, but this is unavoidable since selecting jets based on the number of scatterings means selecting jets with a certain shape and fragmentation pattern.

We here fix the long axis of the brick to \unit[1]{fm} and the temperature of the gluons to \unit[200]{MeV} (this affects only the momentum distribution of gluons making up the background medium). The density, screening mass and eccentricity are varied.

\section{Results}

\begin{figure}[!ht]
    \centering
    \subfigure[]{
        \label{RAANB}
        \includegraphics[width=.45\textwidth]{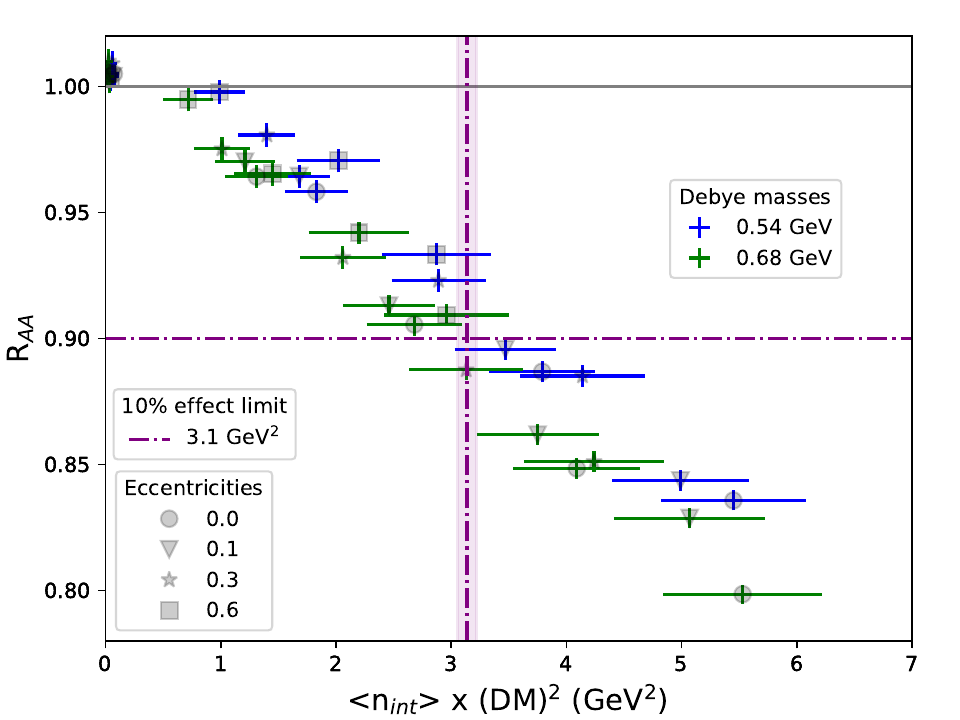}
    } \quad
    \subfigure[]{
        \label{v2NB}
        \includegraphics[width=.45\textwidth]{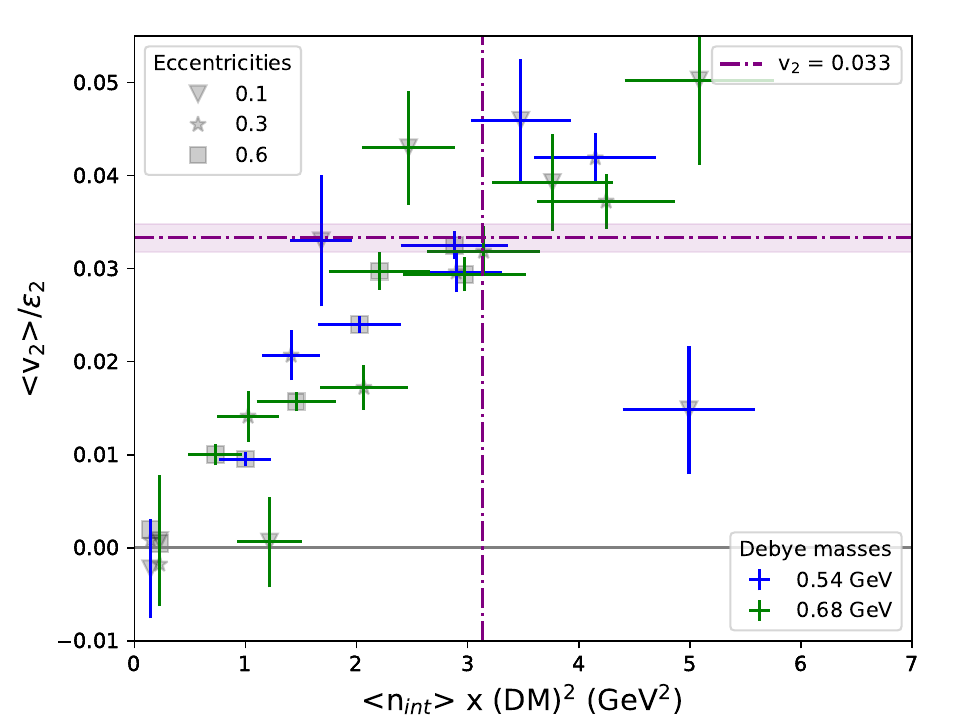}
    }
    \caption{\subref{RAANB} \RAA and \subref{v2NB} \vtwo results also as a function of \avgt.}
    \label{RAAv2}
\end{figure}

We proceed with studying the \RAA dependence on the number of interactions. However, instead of plotting the results as a function of the number of jet-medium interactions, we plot them against the average number of interactions per parton ($\langle\text{n}_{\text{int}}\rangle$) times the square of the Debye mass (DM), which is a proxy for the average momentum squared exchanged per parton between jet and medium. This is done because the DM controls how hard the interactions are, which strongly affects the energy loss per interaction and, consequently, the \RAA.\footnote{{In a single scattering the energy loss of the energetic parton scales approximately as (DM)$^2$.}} To demonstrate the approximate scaling with \avgt, we calculate \RAA and \vtwo as a function of \avgt for two different values of the DM, as seen in figure~\ref{RAAv2}. The number of interactions per parton is calculated taking into account the entire (splitting) history of the partons. At the end of the partonic phase we find all partons present at that stage. We then follow each parton backwards through the evolution counting the number of scatterings. When a splitting is reached we continue with the mother parton. In this way all scatterings in the history of the parton are counted. The resulting distributions of number of scatterings are shown in figure~\ref{fig:distributionNint}.

Figure~\ref{RAAv2} shows that both \RAA and \vtwo scale almost linearly with \avgt. The dot dashed lines in figure \ref{RAANB} show the point at which the \RAA goes below 0.9, which is around the value of \RAA that can be reliably measured. We extract the value of \avgt at which this happens from a linear fit to the \RAA values and obtain 3.1(1) \GeV$^2$, corresponding to 12 interactions per parton in the case of DM=\unit[0.55]{GeV} and around 8 interactions for DM=\unit[0.68]{GeV}. These numbers correspond to a total of around 100 to 150 interactions in the di-jet event. This is counting all scatterings of partons belonging to the hard partonic system, including those of partons that end up outside the reconstructed jets. 

On the other hand, figure \ref{v2NB} shows the result of \vtwo divided by the eccentricity ($\epsilon_2$) of the brick (we find that \vtwo is proportional to $\epsilon_2$). The eccentricity is calculated from the position of the scatterings using $\epsilon_2=<y^2-x^2>/<x^2+y^2>$. In figure~\ref{v2NB}, the dot dashed lines show the value of \vtwo at the same point where a 10\% effect in \RAA was observed in figure \ref{RAANB}. That is, for \avgt = $\unit[3.1]{GeV^2}$, \vtwo$/\epsilon_2 = 0.033(1)$, which corresponds to \vtwo = 0.0099(3) for an intermediate eccentricity of 0.3.
It should be noted that, when computing \vtwo in the way described in section \ref{sec:small systems}, i.e., relative to the symmetry plane, it should be considered as an underestimate of the values obtained in experiments from two particle correlations. The latter is generally larger than the former, for instance, because it responds differently to fluctuations. We, thus, do not see our results as a contradiction to the measurements. Nonetheless, these results indicate that once the system has interacted enough to produce a \vtwo of high $p_\perp$ particles, \RAA should also show substantial suppression.

\begin{figure}[!ht]
    \centering
    \subfigure[]{
        \label{RAAvsL}
        \includegraphics[angle=-90,width=.45\textwidth]{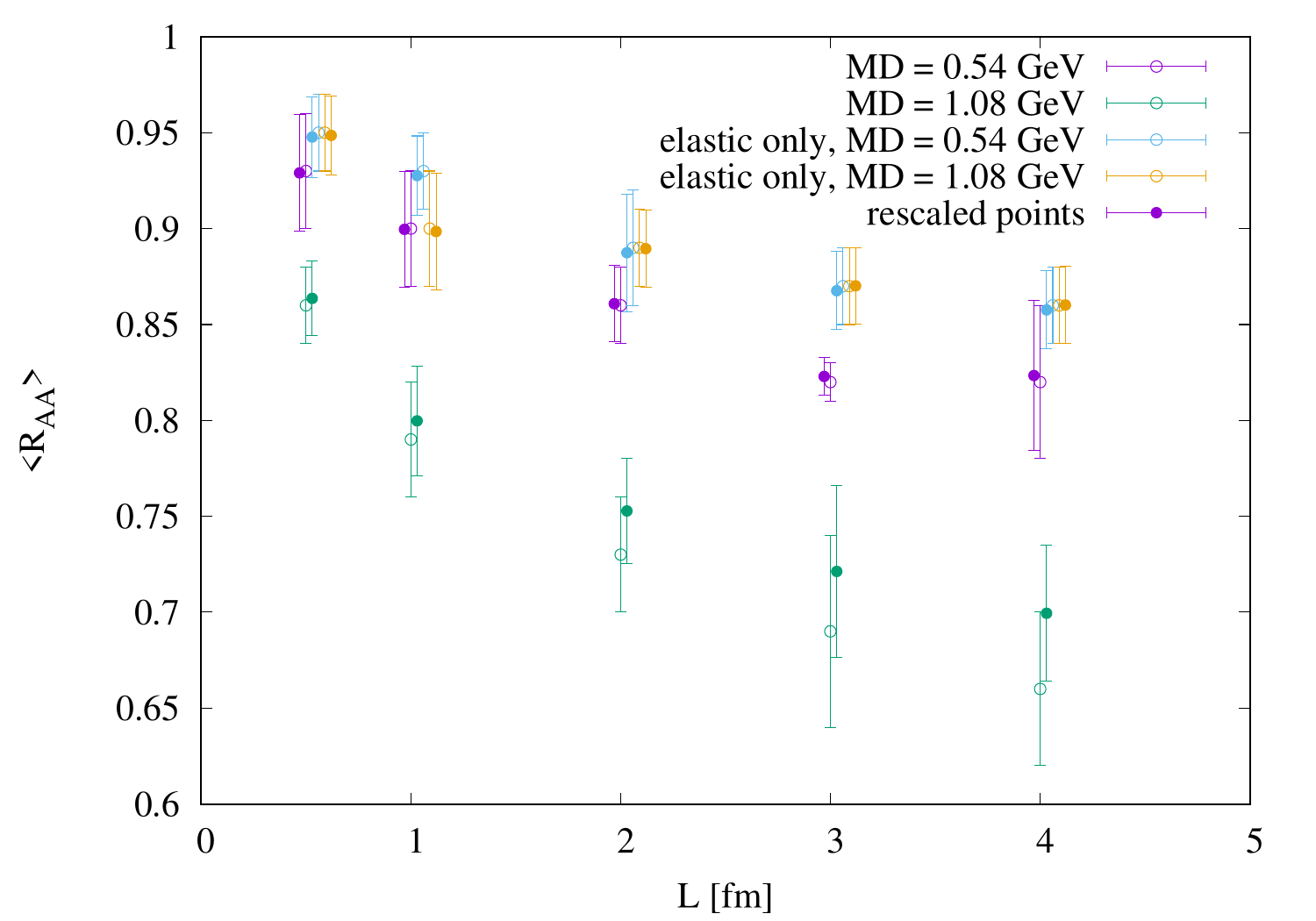}
    } \quad
    \subfigure[]{
        \label{numsplits}
        \includegraphics[angle=-90,width=.45\textwidth]{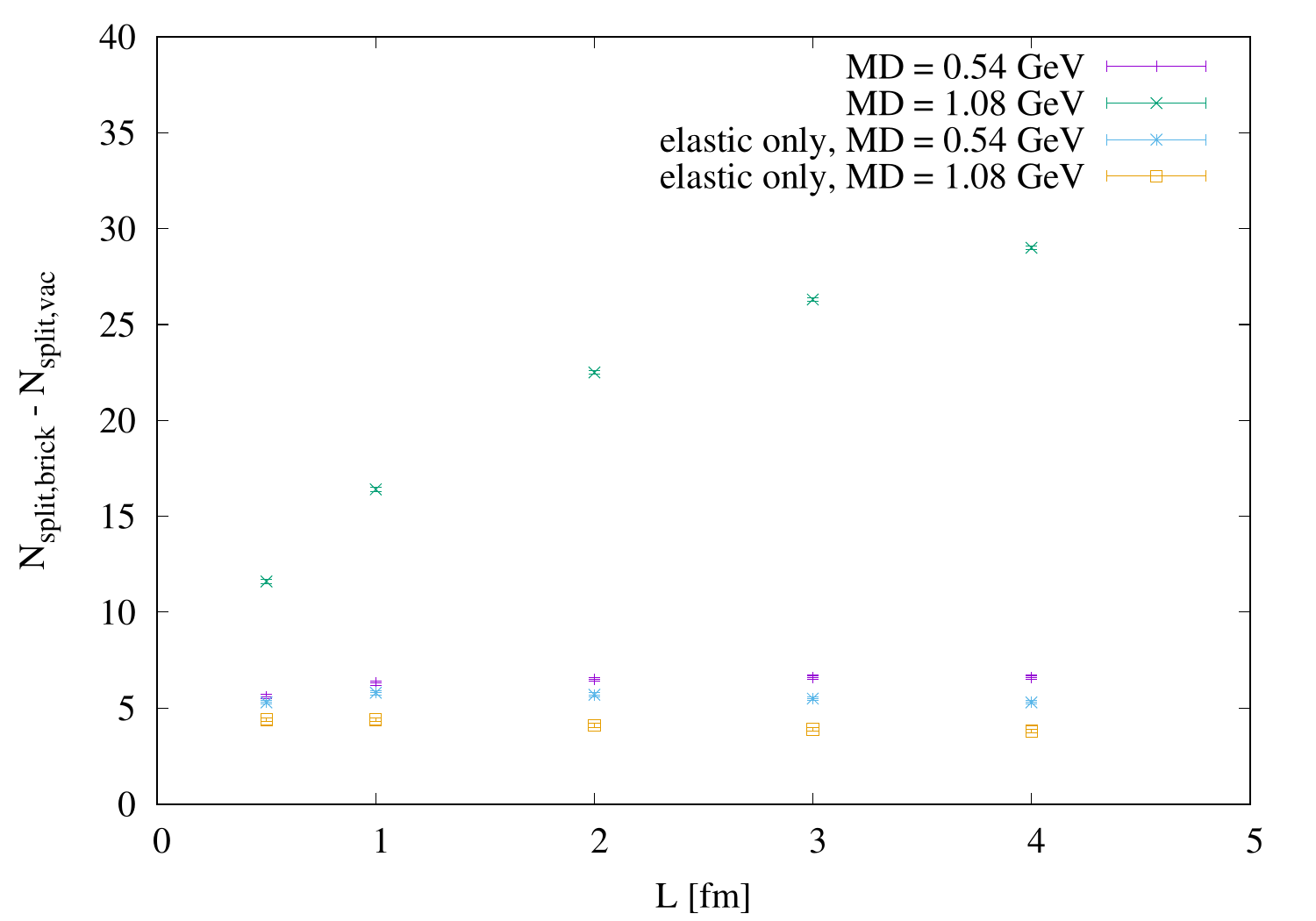}
    }
    \caption{\subref{RAAvsL} \RAA as a function of brick radius for points with similar \avgt. Since the points have slightly different \avgt we have rescaled \RAA to a common \avgt=$\unit[3.5]{GeV^2}$ assuming a linear dependence of \RAA on \avgt. \subref{numsplits} Difference in number of splittings between the medium simulations and the corresponding vacuum corresponding to the points in \subref{RAANB}.}
    \label{scalingcheck}
\end{figure}

We also verified the range of validity of the linear scaling with \avgt. Figure~\ref{RAAvsL} shows the \RAA as a function of brick radius for different parameter settings with the same \avgt. It becomes clear that, as the size of the medium increases, the same value of \avgt produces more suppression when compared with the smaller medium. This is most pronounced for higher values of the Debye mass. In \textsc{Jewel} the amount of medium induced radiation depends rather strongly on the Debye mass, because it is much easier for harder scatterings to initiate medium induced emissions. It is also easier for a scattering to induce an emission at later times, when emissions from the parton shower of the initial hard scattering tend to have longer formation times. This leads to an increase of medium induced emissions per scattering as a function of system size as seen in figure~\ref{numsplits}. Thus, when the inelastic interactions are turned off (i.e. there is only elastic scattering), this suppression is not as important. This indicates  that the breaking of this scaling is largely due to a sizable contribution from inelastic interactions. It is worth noting that, in the medium, angular ordering is not imposed between two splittings if an interaction has taken place between them. Therefore, even the points with only elastic interactions are expected to have more splittings than the jets in vacuum. As for the remaining difference, it is accounted for by medium induced emissions. Therefore, the plots in figure \ref{numsplits} indicate that inelastic energy loss does not follow the scaling, and, in small systems, where the scaling approximately holds, energy loss is mainly driven by elastic interactions. Figure~\ref{scalingcheck} also shows that, while inelastic energy loss is an important factor for departures from scaling, there is an effect even in the absence of inelastic scattering. This is because, even for elastic scattering, the time at which a scattering occurs is not completely irrelevant (as assumed by the scaling). At early times the partons did not have time to radiate much yet and therefore the average parton energy is higher than at later times. A higher parton energy implies a smaller energy loss for the same momentum transfer and an early scattering thus leads to a somewhat smaller energy loss than a later one. Also, early scatterings contribute less to broadening, which transports energy out of the jet cone. This is so because a scattering of a parton early in the branching history tends to deflect the whole jet rather than individual constituents of the jet.

To confirm that the earlier scatterings contribute less to \RAA than the later ones, we have taken the spherical brick with radius of 1 fm and split it into two: one smaller spherical brick of radius 0.5 fm and one hollow spherical brick of radius 1.0 fm which has no medium within the first 0.5 fm radius. Both of them have the same density (0.1 or $\unit[0.2]{GeV^{-3}}$) and Debye mass (\unit[0.68]{GeV}). The results are shown in Table \ref{tab:hollowbrick} and one can see that, for very close values of \avgt, the later interactions lead to a stronger suppression.

\begin{table}[!ht]
\centering
\begin{tabular}{|l|l|l|l|l|}
\hline
\multirow{1}{*}{Density} & \multirow{1}{*}{R$_\text{in}$} & \multirow{1}{*}{R$_\text{out}$} & \multirow{1}{*}{\avgt} & \multirow{1}{*}{\RAA} \\
(GeV$^{-3}$) & (fm) & (fm) & (GeV$^2$) &  \\ \hline
0.1 & 0.0  & 0.5 & 1.6(3) & 0.974(5) \\ \hline
0.1 & 0.5  & 1.0 & 1.7(3) & 0.926(4) \\ \hline
0.2 & 0.0  & 0.5 & 3.4(5) & 0.916(4) \\ \hline
0.2 & 0.5  & 1.0 & 3.6(5) & 0.841(4) \\ \hline
\end{tabular}
\caption{\avgt and \RAA in events where there is a medium between a radius $R_{\text{in}}$ and a radius $R_{\text{out}}$.}
\label{tab:hollowbrick}
\end{table}

\begin{figure}[!ht]
    \centering
    \includegraphics[width=.5\textwidth]{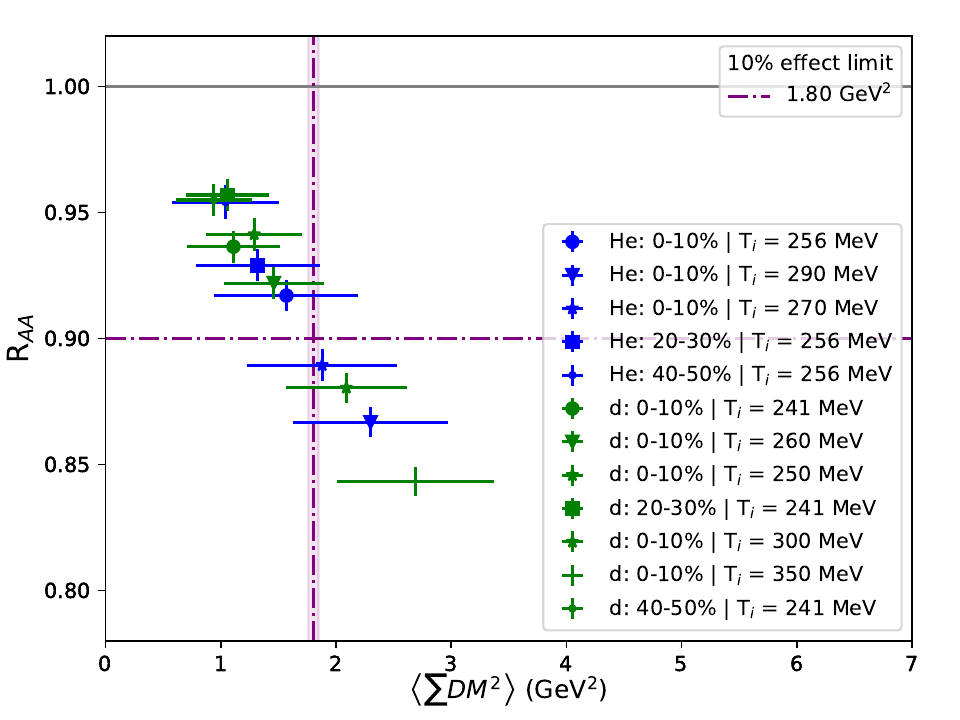}
    \caption{\RAA as a function of $\left< \sum \text{DM}^2 \right>$ for the expanding medium, where DM is allowed to vary and $\left< \sum \text{DM}^2 \right>$ stands for the average sum of DM$^2$ for each interaction per parton.}
    \label{simplemedium}
\end{figure}

Finally, we also generated events using a somewhat more realistic medium model, which accounts for longitudinal expansion and has a temperature profile. This is the simplistic medium model that comes with \textsc{Jewel}~\cite{Zapp:2012ak}. We want to stress that this is by no means a realistic model, and, in particular, it is expected to perform poorly for small systems~\cite{Zapp:2013zya}. The motivation for using it here is rather to have a model that is qualitatively different from the brick medium. We generated events for HeHe and deuteron-deuteron collisions, which yield values of \avgt that are similar to those studied with the brick medium. The initial temperatures were obtained by scaling the energy densities from T$_\text{R}$ENTo~\cite{Moreland:2014oya}, except for the higher values for the 0-10\% centrality classes, which were added to obtain higher values of \avgt. For the initialisation time $\tau_i=\unit[0.4]{fm}$ was used in all cases. It should be noted that now the screening mass varies with temperature throughout the evolution and therefore instead of \avgt we compute for each parton the sum of the actual values of the squared screening mass for each scattering, denoted as $\left< \sum \text{DM}^2 \right>$ (for the brick medium \avgt $ = \left< \sum \text{DM}^2 \right>$). Figure~\ref{simplemedium} shows that with the expanding medium \RAA falls off more steeply with $\left< \sum \text{DM}^2 \right>$ than with the brick medium. This divergence from the result in figure~\ref{RAANB} could be due to the fact that the medium model used was not tailored for application in small systems. In particular, it has a linear increase of the temperature until the initial proper time $\tau_i$ (after which longitudinal expansion leads to a decreasing temperature). For the results shown here $\tau_i$ is a large fraction of the total size and lifetime of a small collision system. Most scatterings, thus, occur at relatively late times and figure~\ref{scalingcheck} indicates that this is expected to lead to a stronger suppression. Nonetheless, figure \ref{simplemedium} also shows a scaling of the \RAA with $\left< \sum \text{DM}^2 \right>$, thus supporting the findings with the brick.

\section{Conclusions}

We have studied how the number of jet-medium interactions affects different jet quenching observables. The first important observation is that the result is largely dependent on the screening mass of the medium and, in fact, \RAA and \vtwo scale approximately with \avgt. Figure~\ref{RAAv2} shows how these observables depend on \avgt and that at $\unit[3.1(1)]{GeV^2}$ a 10\% effect is observed in \RAA. At the same point, and for the intermediate eccentricity studied ($\epsilon_2 = 0.3$), \vtwo = 0.0099(3). This value of \avgt corresponds to a number between 8 and 12 interactions per parton in the range of Debye masses studied. This means that, in total, around 100 -- 150 interactions between the fragmenting partonic system (per event for pure QCD hard scatterings) and medium particles are required to obtain an observable \RAA in small systems. But we have also studied the validity of this scaling behavior and figure \ref{scalingcheck} shows that it breaks down for larger system sizes. It also shows that, when the system size is small, the energy loss happens mainly through elastic scatterings. The reasons for departure from scaling are radiative energy loss, which does not follow the scaling and in \textsc{Jewel} is more important at later times, and the fact that elastic scattering is more effective for jet quenching when it occurs at later times. Finally, we looked at a more realistic medium model and saw that \RAA shows a stronger dependence on $\left< \sum \text{DM}^2 \right>$ than the brick medium. This shows that there is a scaling behavior of \RAA and \vtwo with the number of scatterings and the square of the Debye mass, but the size of the system and the spatio-temporal distribution of scatterings also plays a role.

\section*{Acknowledgments}

We would like to thank Alice Ohlson for enlightening discussions.
This study is part of a project that has received funding from the European Research Council (ERC) under the European Union's Horizon 2020 research and innovation programme  (Grant agreement No. 803183, collectiveQCD). JGM was partly supported ERC under the European Union's Horizon 2020 research and innovation programme  (Grant agreement No. 835105, YoctoLHC).

\bibliographystyle{unsrt} 
\bibliography{refs}

\end{document}